\documentstyle[11pt,paspconf,epsf]{article}

\begin{document}

\title{Brown Dwarfs: From Mythical to Ubiquitous}

\author{James Liebert}
\affil{Steward Observatory, University of Arizona, Tucson, AZ 85721}

\begin{abstract}

Astrophysical objects below the stellar mass limit but well above the
mass of Jupiter eluded discovery for nearly three decades after Kumar
first proposed their existence, and for two decades after Tarter
proposed the name ``brown dwarfs.''  The first unambiguous discoveries
of planetary (51~Peg~B) and brown dwarf (Gliese~229B) companions
occurred about three years ago.  Yet while extrasolar planets are now
being discovered at a breathtaking rate, brown dwarf companions to
ordinary stars are apparently rare; likewise, imaging surveys show
that GL~229B is still unique as a distant companion to a low mass
star.

On the other hand, the deep imaging studies of the Pleiades and
several imbedded young clusters show that the mass function (ie. of
single objects) extends in substantial numbers down to at least 40
Jupiter masses.  The high mass/ stellar density Orion Nebula Cluster
may have relatively fewer low mass objects.  

In the field of the solar neighborhood, the infrared sky surveys DENIS
and especially 2MASS show that brown dwarfs, certified by the lithium
test, exist in significant numbers.  These appear to include most of
the newly-defined spectroscopic class of L dwarfs; these objects are
cooler than, and with different atomic an molecular absorption
features than the late M dwarfs.  If the first 1\% of sky analyzed is
not atypical, over a thousand L dwarfs should be detected in the 2MASS
survey.

\end{abstract}

\section{A Checkered History}

Kumar (1963) first showed that the minimum mass for stellar objects to
fully stabilize themselves by hydrogen-burning is 0.07$M_\odot$ (or a
bit higher).  The calculations of Grossman, \& Graboske (1974)
suggested a terminus of 0.08$M_\odot$, and the best calculations today
predict that the boundary falls between the above two values (for
solar composition). Tarter (1975) first proposed the term ``brown
dwarfs'' for objects below this main sequence limit, but substantially
larger than the mass of Jupiter (M$_J$).  Her thesis advanced the
hypothesis that such objects could provide the ``missing mass'' known
dynamically to exist in galaxies and clusters of galaxies.  Other
suggested designations for these hypothetical objects, such as black
or infrared dwarfs, were also proposed.

By 1986, the time of the first conference dedicated to this subject
(Kafatos, Harrington \& Maran 1986), Tarter's label was widely
accepted (Tarter 1986); however, not a single unambiguous case of a
substellar object of this type was known.  In particular, the object
featured on the cover of that particular conference proceedings turned
out to be perplexingly mythical.

The next several years featured many claims of the discovery of brown
dwarfs (hereafter BDs) and extrasolar planets -- both in the field, in
clusters, and as companions to known stars.  However, these cases
proved unsustainable or ambiguous.  A big part of the problem was that
nobody knew where the terminus really lay in an observational H-R
diagram.  Fainter and fainter objects were being found that appeared
to have temperatures and luminosities that might be interpreted as an
extension of the main sequence, both in the field and in clusters
(especially the Hyades and Pleiades).  The most extreme case by the
late 1980s was GD~165B, found by Becklin \& Zuckerman (1989) with an
estimated M$_{bol}$ = 14.99.  Even this object could be interpreted
barely as stellar, if interior models of high opacity were adopted.

In particular, D'Antona \& Mazzitelli (1985, DM85) predicted that the
main sequence could stretch a factor of ten lower in luminosity in
comparison with earlier models, and pointed out the existence of what
they called objects of ``transition mass.''  Their calculations showed
that, near 0.07$M_\odot$, the configuration initiates proton-proton
burning, for a period of time which decreases with mass.  However,
since the pressure support generated from the nuclear reactions never
quite achieves 100\% of what is needed to prevent a slow contraction
of the star, the increasing densities bring an increasing degenerate
electron pressure.  The onset of energy transport by conduction
results in the eventual decrease of the central temperature, which
shuts off the nuclear reactions -- but, for their 0.075$M_\odot$
model, only after 10$^{10}$ years!

The results of DM85 meant that the main sequence terminus itself, for
solar composition at least, is quite ``fuzzy.'' That is, the criteria
for defining a BD become fuzzy as well.  Should the 0.075$M_\odot$
model of DM85 be called a brown dwarf or a star?  Moreover, even
objects with a limited or no hydrogen-burning phase traverse paths
close to the zero age main sequence, spending substantial time in
gravitational contraction to their final radii.  Whether stellar or
substellar, the mass assignable to a given luminosity is therefore
very sensitive to the age.  The small differences in T$_{eff}$ at a
given luminosity are well below the accuracies of current models in
fitting observations, for a wide range of mass.

The above quandary obviously underscores the value of searches for
substellar objects in the nearest clusters.  The age of the cluster is
known (with possible caveats), and the young brown dwarfs are more
luminous anyway than older counterparts.  However, for those of us
searching in the field of the local disk, the burden of proof is very
high -- it is necessary to determine or bound the mass of the object
independently of the HR Diagram.  One long sought solution is to find
an astrometric or spectroscopic binary with a substellar component.
Indeed, Latham et al. (1989) detected a radial velocity companion to
the F9~V star HD~114762 with a mass function implying
M$_2$~sin~i$_{orb}$ = 0.011M$_\odot$.  However, Cochran, Hatzes \&
Hancock (1991) presented strong evidence based on studying the line
profiles of the star that our view of this system is close to pole-on,
underscoring the possibility that the companion need not be
substellar.

Fortunately, an easier solution has emerged, pioneered by Rafael
Rebolo, Martin \& Magazzu (1992): This is the so-called ``lithium
test'' -- the detection of Li in the spectrum for a configuration
predicted to be completely convective throughout.  Since Li is
destroyed at central temperatures of hundreds of thousands of degrees,
its survival means that hydrogen burning (at over a million degrees)
is not taking place.  However, even the Li~I resonance transitions in
the red spectrum generally are weak lines of an element of low
abundance.  Thus high resolution spectra are generally required, of
very cool objects having precious little red flux.

By the mid-1990s, even the cluster searches had failed to uncover
ironclad candidates, and neither the appropriate binary nor a
candidate showing lithium had turned up in the field.  Then, two
nearly-simultaneous discoveries were announced in the same volume of
{\it Nature}.  These were (1) the first extrasolar planet, 51~Cyg~B,
from radial velocity searches (Mayor \& Queloz 1995), and (2) the
first unambiguous brown dwarf, Gliese 229B (Nakajima et al. 1995). 
In three short years since then, a plethora of extrasolar planets 
and BDs have been unveiled. 

The BD discoveries can be divided into those found (1) as companions
to known nearby stars, (2) as members of nearby star clusters,
including very young objects in giant molecular clouds, and (3) as
field objects in the solar neighborhood.  For reasons of space and the
degree of my own involvment, a relatively greater emphasis is placed 
on the discovery of field objects, especially results from the Two 
Micron All Sky Survey (2MASS).

\begin{figure}
\centerline{\epsfxsize=3.5in\epsfbox{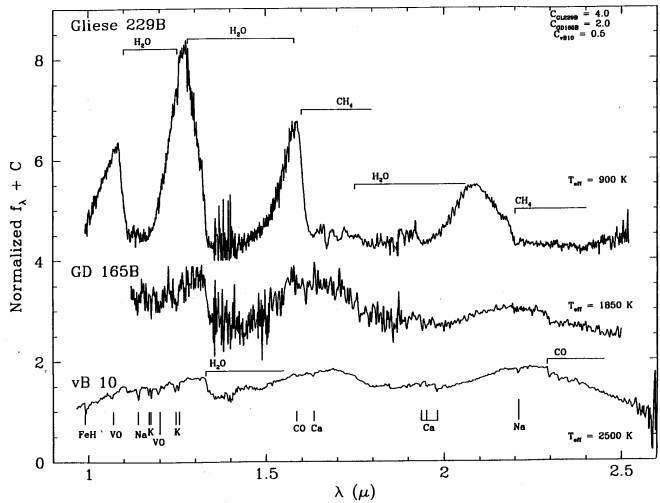}}
%\vspace{7.4truecm}
%\special{psfile=fig1.ps voffset=-10 hoffset=80
%vscale=33 hscale=33 angle=0}
\caption{The near IR spectrum of GL~229B compared with spectra of vB10
(M8~V) and GD~165B (L2~V, see section 4) .  The great strengths of the 
CH$_4$ and H$_2$O bands in GL~229B give it a blue J-K color. From
Geballe et al. 1996, Jones et al. 1994, and Oppenheimer et al. 1998b.}
\label{fig-1}
\end{figure}

\section{Brown Dwarf Companions to Nearby Solar-Type Stars}

At the time of this writing, GL~229B has remained unique, as the only
directly-detected brown dwarf companion with T$_{eff}$ near 1,000~K
and methane in its spectrum.  In Fig.~1 is shown the near-IR spectrum
of this object, showing deep absorption features due to H$_2$O and
CH$_4$, compared with spectra of GD~165B (an L dwarf) and the benchmark
late M dwarf vB~10 (from Oppenheimer et al. 1998b). The need to explain
the unusual atmospheric composition of GL~229B, the role of chemical
equilibrium of gas species and various kinds of dust, mixing
processes, the likely spatial and time variability (``weather'') --
make it a new astronomical field all to itself.  Further discussion is
beyond the scope of this paper, but the reader is referred to
Oppenheimer et al. (1998b) and to Burrows \& Sharp (1998) for,
respectively, an observational and theoretical sampling.

In contrast, the inferred discoveries of extrasolar planetary
companions to nearby main sequence stars by Butler \& Marcy (1996),
the Mayor group, and others have continued at a fast and furious pace.
At first it appeared also that the radial velocity companions included
a substantial fraction of BDs -- defined (somewhat arbitrarily) as
0.01M$_\odot$ and larger -- generally in quite eccentric orbits.
Recently, however, Mayor et al. (1998) reported that orbital
inclination determinations based on HIPPARCOS astrometry show that
several of the companions previously believed to be substellar have
orbital inclinations high enough to lie near or above the stellar mass
limit.  A few known cases remaining with lower-limit masses well below
the stellar limit do {\it not} have accurate orbital inclination
determinations.  The possibility exists at the time of this writing
that the fairly close substellar companions amenable to detection from
radial velocity surveys include brown dwarfs only rarely, if at all.

In support of this conclusion may also be the imaging survey of
Oppenheimer et al. (1998a); following their early discovery of GL~229B,
they have up to now reported no new substellar companions to the over
100 solar neighbors in their survey.  Thus, it now appears from both
types of observations that BDs orbitting solar-type stars are rare.

\section{Brown Dwarfs in Star Clusters}

\subsection{The Pleiades} 

The studies of star clusters began bearing real fruit -- verifiable
brown dwarf members -- also in the last few years.  This followed a
series of false starts and ambiguous results, many involving the
Hyades cluster and the Taurus clouds.  The need to survey large areas
with difficult background fields have plagued the studies of these two
stellar aggregations.  The Pleiades cluster is younger but more
distant than the Hyades.  Coincidentally, the two effects nearly
offset, so that the BDs were predicted to have about the same
luminosities as the Hyades counterparts.  Moreover, the Pleiades is a
much more compact cluster, with a less difficult background field.  It
is not surprising, therefore, that it has now yielded some of the most
solid BD candidates.  Here we sketch incompletely some major findings.

The first important Pleiades candidate, PPL~15, emerged from a deep
CCD survey of Stauffer, Hamilton and Probst (1994).  Basri, Marcy and
Graham (1996) detected lithium for the first time in this cluster
candidate, but even the apparent passing of the lithium test left an
ambiguous situation.  The age of the Pleiades has been estimated
variously between 70 Myrs for upper main sequence interiors with no 
convective core overshooting, to 120 Myrs if overshooting adds more 
hydrogen fuel to the core.  With the former value, the luminosity of
PPL~15 corresponds to a substellar mass.  However, the Li detection
was weak enough that the authors found that it had been partially
depleted.  The most consistent picture at the time was provided by 
concluding that the age was closer to 120 Myrs, and the mass right
at the hydrogen-burning mass limit.  In the last few years, lower
luminosity BD candidates have been found, though we shall return to
the saga of PPL~15.

Deeper CCD surveys performed primarily in the Canaries by European
investigators found two important brown dwarf candidates, Teide~1
and Calar~3.  Both have spectral types of M8 (Martin, Rebolo \&
Zapatero Osorio 1996), yielding masses of 0.055$\pm$0.015M$_\odot$
(Rebolo et al. 1996), assuming the 120 Myr age. Most importantly, high
resolution spectra show that these have kinematics consistent with
cluster membership, and detections of apparently-undepleted lithium.
In the last two years, even deeper surveys have been performed which
emphasize redder CCD bandpasses like I and Z (cf.  Zapatero Osorio et
al. 1997) and resulting in the discoveries of the even-fainter
``Roque'' candidates (named for the unique rock formation at the
summit of La Palma).  The faintest of these appears to cross into the
L spectral class discussed in Sect.~4.2 with a corresponding mass 
below 40 M$_J$ (Martin et al. 1998).

At the time of this writing, there are of the order 50 photometric BD
candidates from these studies.  The inferred luminosity function of
the Spanish/European groups suggests that the number of BDs in the
cluster down to 30-40 M$_J$ could be of the order of a few hundred, a
number density comparable to that of stars. The mass contribution
would of course be considerably less, though the slope of the mass
function in logarithmic units appears to be flat to slightly negative
(slightly more objects with decreasing mass).  Moreover, a consistent
lithium boundary line -- below which lithium always appears, above
which it does not -- lies near the hydrogen-burning mass limit for the
Pleiades age.
 
Finally, PPL~15 was found to be the first binary brown dwarf, using
$HIRES$ on Keck~I (Basri \& Martin 1998).  The excessive luminosity is
explained by two components, each of mass near 0.065M$_\odot$,
contributing roughly equally to the total light.  The period is 5.8
days, at a separation of a few R$_\odot$, and an orbital eccentricity
near 0.5.

\subsection{Imbedded Clusters} 

Younger, nearby star-formation regions offer more luminous brown
dwarfs, though one usually has to cope with dust extinction by
working at infrared wavelengths.  The following examples offer only an
incomplete summary of studies of the low end of the mass function in
several imbedded regions.  The $\rho$~Ophiuchis cluster is one of the
nearest but most imbedded (Comeron et al. 1993; Luhman \& Rieke, in
preparation, see also Strom, Kepner, \& Strom 1995).  Others include
NGC~2024 (Comeron et al. 1996), L1495E in Orion (Luhman \& Rieke
1998), and IC348 (Luhman et al. 1998).  There is little doubt that
these studies find candidates extending well into the substellar mass
range, perhaps to the 30-40 M$_J$ reached in the Pleiades.  These
authors generally conclude that the log mass function -- again in
apparent agreement with the Pleiades -- is flat, or with a small
negative slope, for the clusters they have studied.  On the other
hand, Hillebrand (1997) in a comprehensive study of the rich Orion
Nebula Cluster (ONC) finds that its mass function peaks near
0.2M$_\odot$ and falls rapidly towards lower masses.  This ONC region
of higher stellar density with many massive stars may thus be a less
hospitable environment for the formation of very low mass stars and
brown dwarfs.  However, the ONC is also generally more distant than
the other clusters, so that survey completeness below the stellar mass
limit may be a more difficult issue.

\begin{figure}
\centerline{\epsfxsize=3in\epsfbox{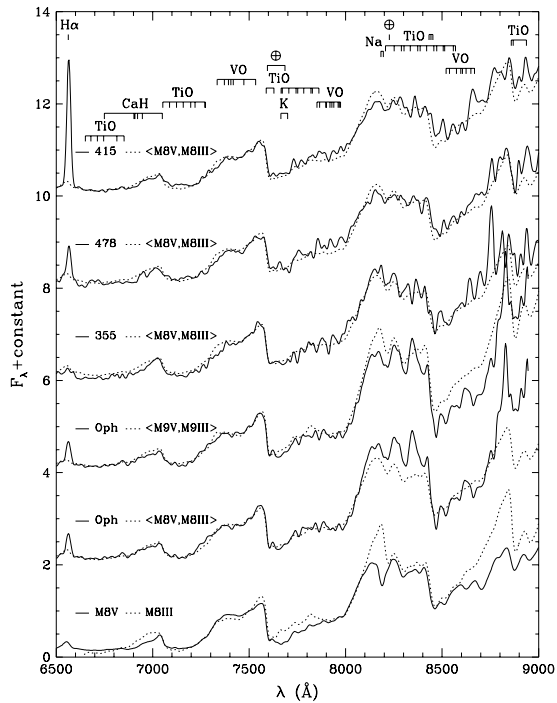}}
%\vspace{8.5truecm}
%\special{psfile=fig2.ps voffset=-10 hoffset=80
%vscale=33 hscale=33 angle=0}
\caption{Four brown dwarfs well below the stellar mass limit, found by
Luhman et al. (1998) in IC348 and $\rho$~Oph. The three latest sources
observed in IC~348 (415=M7.5, 478=M7.5, 355=M8) and $\rho$~Oph
162349.8-242601 (M8.5) (solid lines) are plotted with averages of
standard M8 and M9 dwarfs and giants (dotted lines).  Features which
are sensitive to surface gravity are apparent in the comparison of the
M8~V and M8~III spectra.  All spectra are normalized at 7500~\AA.}
\label{fig-2}
\end{figure}

That the above photometric studies are finding legitimate BD members
of the clusters is again demonstrated by spectra.  In Fig.~2 several
spectra are shown (courtesy of Kevin Luhman), which demonstrate that
these are young BD members of low mass.  Their spectra contrast with
those of GL~229B and the older, field BD candidates discussed in the
next section.  The young imbedded cluster candidates are both warmer
and more luminous: an M7-M9 spectrum can come from a young object of
several tens of Jupiter masses.  Moreover, these differ from the
spectra of dwarfs.  The gravity-dependent features indicate objects
intermediate in luminosity between dwarfs and giants.  Finally, the
identification of legitimate BDs is important in order to study
whether proto-planetary disks can form around such objects, analogous
to what produced the Galilean moon system around Jupiter, and whether
these low mass objects show T~Tauri-like activity, or far-infrared
excesses.

\section{The First Field Brown Dwarfs}

Both the results from the clusters discussed above and the discoveries
of GD~165B and GL~229B many tens of a.u. from their companion stars
indicated that brown dwarfs could be found in the solar neighborhood.
The problem is that such objects might represent a wide range of age
in the Galactic disk, and even the distances would be initially
unknown.  On the other hand, the lithium test might be applied, or,
for extremely cool cases, the methane test.  For GD~165B, even an
excellent Keck~II spectrum lacks sufficient flux at 6700\AA\ to test
for lithium.  The new study by Kirkpatrick et al. (1998a) makes the
case that it is a brown dwarf, although the minimum age assignable
from the cooling time of the white dwarf companion suggests that it is
undergoing at least a long transitory phase of nuclear-burning.  (I
reiterate that the definition of a ``brown dwarf'' is ambiguous!)
However, in the last few years some cool field objects have been found
which pass the lithium test, which means that the central temperatures 
are low enough that any hydrogen-burning phase would have been brief.

From UK and ESO Schmidt plates, Thackrah, Jones \& Hawkins (1997)
found an M6 dwarf which shows a distinct Li~I 6707\AA\ detection.
Depending on the age, T$_{eff}$, implied Li abundance, and the
uncertainties due to models, it still appears to straddle the stellar
boundary.  A second, much cooler object is Kelu~1, found in the
Chilean proper motion survey by Ruiz, Leggett \& Allard (1997).  The
model atmospheres generated by Allard suggest that T$_{eff}$ is about
1,900~K.  Its spectrum places this object among the L dwarfs discussed
in Sect.~4.2.  The presence of Li makes it probable that this object
is substellar.  A third case has been found from an old Luyten proper
motion catalog, LP~944-20 (Tinney 1998).  This object has been
assigned spectral type M9~V by Kirkpatrick, Henry \& Simons (1995),
making the T$_{eff}$ intermediate between the first two objects
discussed above.  Tinney (1998) argues that the mass is near
0.06~M$_\odot$, and the age 475-700~Myr.  Again, the sequence of Li
observations in the Pleiades -- where ages and luminosities are known
-- lends credence to the model-dependent interpretations of the field
objects.

\subsection{The DENIS and 2MASS Surveys}

Finding the nearest, ultracool solar neighbors is extremely valuable
since these will be the brightest such objects, and most amenable to
detailed followup studies.  Fortunately, within the last few years,
two groups have begun the first substantial sky surveys at
near-infrared wavelengths.  The DEep Near Infrared Sky survey (DENIS)
was started in January 1996 by a consortium of European investigators
for the southern hemisphere.  The Two Micron All Sky Survey (2MASS)
began in May 1997 at Mt. Hopkins for the northern sky, and at CTIO in
February 1998 for the southern sky (Skrutskie et al. 1997).

In the paper by Delfosse et al. (1997), DENIS announced the discovery
of the first brown dwarf candidates from an infrared survey; in a
companion paper, Martin et al. (1997) reported the detection of
lithium in one of these candidates, demonstrating that it is a BD near
60~M$_J$.  The DENIS candidates also showed spectra obviously later
than the end of the defined M sequence (M9/M9.5~V), with qualitatively
different spectral features.  It is in the latter citation above that
a brief proposal first appears in a refereed journal that these
objects be called L dwarfs, following a suggestion by Kirkpatrick
(1997).  We shall discuss the characteristics of these objects below.

In the first 400 square degrees of 2MASS data, plus some other fields
analyzed with protocamera scans, 20 L dwarfs were found (Kirkpatrick
et al. 1998b, hereafter K98).  A total of 25 were known at the time of
the preparation of this paper, including the three DENIS objects,
Kelu~1, and GD~165B.  The 2MASS selection criterion which has a high
degree of success in identifying very late M and L dwarfs is that J-K
$>$ 1.3, while R-K $>$ 5.5, where ``R'' is the photographic red
magnitude from the Palomar Observatory Sky Survey (POSS~1 or 2).

\subsection{The L Dwarfs} 

\begin{figure}
\centerline{\epsfxsize=3in\epsfbox{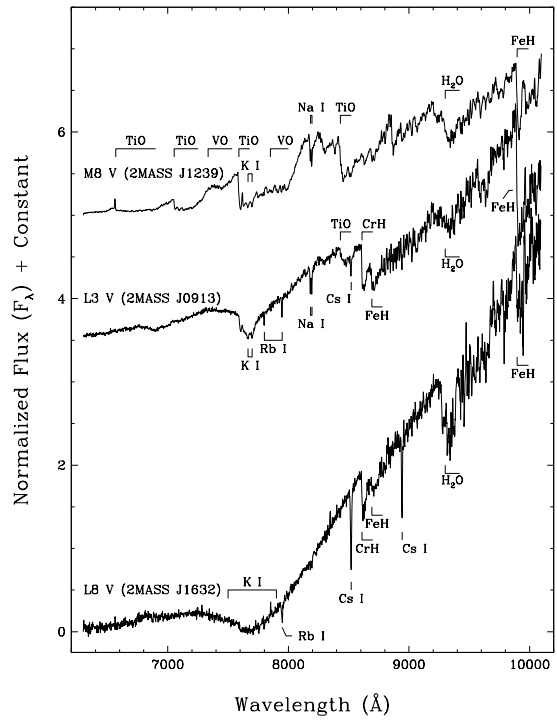}}
%\vspace{8.6truecm}
%\special{psfile=fig3.ps voffset=-10 hoffset=80
%vscale=33 hscale=33 angle=0}
\caption{Red spectra of an M8~V dwarf (top) an L3~V dwarf (middle) and
very late L8~V dwarf (bottom).  Note the dominating strength and width
of the K~I resonance doublet in the L8~V spectrum, and other resonance
lines of rarer alkalis, along with the disappearance of the TiO and VO
bands.  }
\label{fig-3}
\end{figure}

A complete classification system has been developed in K98; some
principal features are summarized here.  In Fig.~3 spectrophotometry
is shown of one late M dwarf, one middle and one very late 2MASS L
dwarf.  These cover ``far red'' wavelengths (6200-10000\AA), and were
obtained with the Keck~II low resolution spectrograph (LRIS).

The strongest features at these wavelengths in late M dwarfs are the
prolific band systems of titanium and vanadium oxides, just as water
and carbon monoxide dominate at longer wavelengths.  So strong are
these that nearby ``pseudo-continuum'' peaks tower high above the
wavelengths where the absorptions are strongest, though at no
wavelength is the true continuum reached.  By the latest M types, 
however, the TiO strengths have peaked out and begun to weaken. 
At the slightly cooler T$_{eff}$ where VO begins to weaken as 
well, this is defined as the beginning of the L sequence (L0~V). 
With progressively increasing L type (and decreasing T$_{eff}$), 
the TiO and VO bands weaken (in the L3~V dwarf of Fig.~3) and then 
disappear entirely (the L8~V object). 

Our understanding of the physics of the above phenomena is as follows:
The M dwarf temperature scale is not known to better than 10\% or so,
but it is believed that below about 2500~K, first the TiO and then VO
molecules in the atmospheres precipitate out as dust grains (Tsuji et
al. 1996ab; Allard 1997; Burrows and Sharp 1998).  By removing these
principal sources of opacity from the atmospheres, dwarfs a few
hundred degrees cooler than late M stars have more transparent
atmospheres at these wavelengths.  Among the remaining absorbers are
hydride bands with limited wavelength coverage.  Otherwise, there is
little continuum opacity, due to the extremely low electron density
(for H$^-$ and H$_2^-$).  Thus, the atomic resonance lines of the
neutral alkalis become quite strong for an abundant species like K~I,
and easily detected for rare species like Rb~I, Cs~I, and if
undepleted, Li~I.  That is, the crucial spectral identifier (Li~I) of
high mass brown dwarfs appears strongly enough at an undepleted
abundance to be detectable on low resolution spectra, rather than
requiring the high resolution needed for M dwarfs -- see Fig.~4,
discussed below.  At late L types, these trends carry to extreme.  As
is evident for the L8~V dwarf in Fig.~3, the K~I doublet with great
width and strength becomes the dominant feature of the red spectrum.
(The Na~I resonance doublet would be even stronger, if our spectra
extended to short enough wavelengths.)

Late L dwarf optical spectra are in one respect analogous to those of
cool white dwarfs, which may also show strong, pressure-broadened
lines.  On the other hand, in the near infrared, the L dwarfs show
very strong CO and H$_2$O, but these are the same features seen in M
dwarfs (see GD~165B in Fig.~1), .  The effective temperatures of the
coolest L dwarfs are not known, but in any case they are not cool
enough for the CO molecule to give way to CH$_4$, as it is predicted
to do near 1500~K (Burrows \& Sharp 1998).  The only known ``methane
dwarf'' remains GL~229B.

The onset of methane in the near-infrared spectrum will reverse the
trend in J-K color with decreasing temperature (Burrows et al. 1997).
GL~229B has a very blue J-K= -0.1. Thus the selection criteria must be
modified to search for objects with blue or neutral J-K color,
increasing enormously the numbers of apparent point sources which
must be screened.  At the time of this writing, it must be concluded
that we have not yet searched effectively for field ``methane
dwarfs.''

The yield of the L dwarfs is substantial enough to conclude that they
are present in significant numbers in the field around the Sun.  It is
elementary to do the math: Some 15 were found in the first 400 square
degrees (1\%) of sky, which means that the entire sky may yield at
least 1,000 objects to the 2MASS survey depth.  (This neglects the
likelihood that the current color selection techniques do not find
100\% of the L dwarfs, not to mention methane dwarfs.)  It is too
early for a first estimate of the space density of substellar objects,
since the distances and luminosities are not yet known (not to mention
the ages and masses).

\begin{figure}
\centerline{\epsfxsize=3in\epsfbox{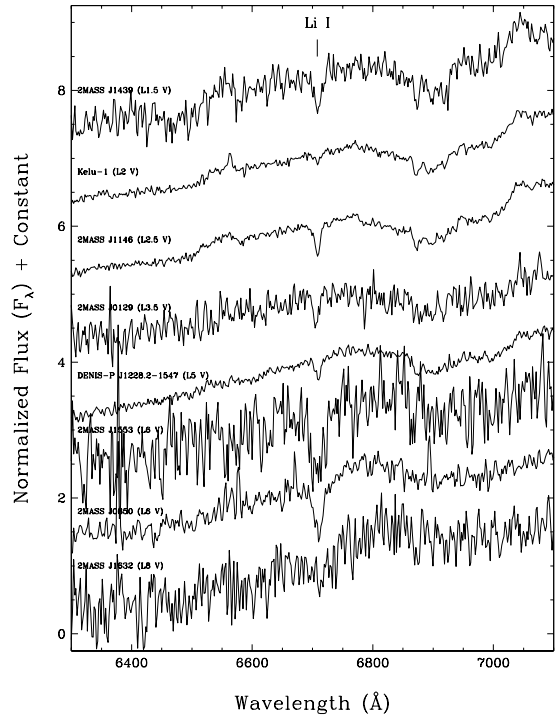}}
%\vspace{8.45truecm}
%\special{psfile=fig4.ps voffset=-10 hoffset=80
%vscale=33 hscale=33 angle=0}
\caption{Spectra of a sequence of L dwarfs, increasing in type from 
to to bottom, centered on the Li~I 6707\AA\ resonance doublet.  The 
line is generally detected, but highly variable in strength.  The  
H$\alpha$ line is also included, but appears convincingly in emission 
only for Kelu~1. }
\label{fig-4}
\end{figure}

We can show that many are substellar, however.  In Fig.~4, a handful
of our objects are illustrated in spectra centered on the Li~I
6707\AA\ resonance doublet.  The candidates observed with Keck~II fall
into three categories: (1) those with Li~I detections at equivalent
widths of several Angstroms, due to the high atmospheric transparency
discussed earlier; (2) those with good enough spectra to show that
Li~I is depleted and undetected; and (3) those too faint for a decent
spectrum to be achieved at these wavelengths, even with a 10 meter
telescope.  The statistics suggest that most of the L dwarfs are
indeed brown dwarfs, but the reader is referred to K98 for
justification of this conclusion.

\acknowledgments

This work was supported by NASA's Jet Propulsion Lab through contract 
number 961040NSF, a core science grant to the Two Micron All Sky 
Survey science team.  I also acknowledge support from the National 
Science Foundation through grant AST~92--17961.

\end{document}